\documentclass[aps,prb,twocolumn,superscriptaddress, show pacs]{revtex4-1} 
\UseRawInputEncoding
\usepackage{bm, amsmath}
\usepackage{graphicx}
\usepackage{color}
\usepackage{epstopdf}
\usepackage{graphicx}
\usepackage{dcolumn}
\usepackage{bm}
\usepackage{subfigure}

\begin{document}

\title{Field-driven metamorphoses of isolated skyrmions within the conical state of cubic helimagnets}

\author{Andrey O. Leonov}
\thanks{leonov@hiroshima-u.ac.jp}
\affiliation{Department of Chemistry, Faculty of Science, Hiroshima University Kagamiyama, Higashi Hiroshima, Hiroshima 739-8526, Japan}
\affiliation{Chirality Research Center, Hiroshima University, Higashi Hiroshima, Hiroshima 739-8526, Japan}
\affiliation{IFW Dresden, Postfach 270016, D-01171 Dresden, Germany} 

\author{Catherine Pappas}
\thanks{c.pappas@tudelft.nl}
\affiliation{Faculty of Applied Sciences, Delft University of Technology, Mekelweg 15, 2629JB Delft, The Netherlands}
\affiliation{Chirality Research Center, Hiroshima University, Higashi Hiroshima, Hiroshima 739-8526, Japan}

\author{Ivan I. Smalyukh}
\thanks{ivan.smalyukh@colorado.edu}
\affiliation{Soft Materials Research Center and Materials Science and Engineering Program,
University of Colorado, Boulder, CO 80309, USA}
\affiliation{Department of Physics and Department of Electrical, Computer and Energy Engineering,
University of Colorado, Boulder, CO 80309, USA}
\affiliation{Renewable and Sustainable Energy Institute, National Renewable Energy Laboratory
and University of Colorado, Boulder, CO 80309, USA} 
\affiliation{Chirality Research Center, Hiroshima University, Higashi Hiroshima, Hiroshima 739-8526, Japan}

\date{\today}


         
\begin{abstract}
{Topologically stable field configurations appear in many fields of physics, from elementary particles to condensed matter and cosmology.
Deep physical relations and common physical features of a large variety of very different solitonic states in these systems arise from the mathematical similarity of phenomenological equations. 
During the last decade, chiral liquid crystals and chiral magnets took on the role of model objects for experimental investigation of topological solitons and understanding of their nonsingular field configurations.
This is directly related to the discovery of particle-like chiral skyrmions, which are also considered as promising ingredients for technological applications. 
Here we introduce a paradigm of facile skyrmionic networks with mutually-orthogonal orientations of constituent isolated skyrmions. 
On the one hand, such networks are envisioned as  a novel concept of spintronic devices based, e.g., on gapless skyrmion motion along each other, and are presumably  responsible for precursor phenomena near the ordering temperatures of bulk cubic helimagnets. 
In particular, we demonstrate an interconversion between mutually orthogonal skyrmions: horizontal skyrmions may swirl  into an intermediate spring-like states and subsequently squeeze into vertical skyrmions with both polarities. 
On the other hand, skyrmion tubes are considered as building blocks for particle-like states with more involved internal structure. 
A family of target-skyrmions, which includes an overlooked so far type with a multiple topological charge, is formed owing to the tendency to minimize the interaction energy between vertical and horizontal skyrmions. 
%
%
The conical phase serves as a suitable background for considered skyrmion evolution. It substantializes the attracting skyrmion-skyrmion interaction in the skyrmionic networks, shapes their internal structure and guides the nucleation processes. 
Alternatively, intricate textural changes of isolated skyrmions result not only in the structural deformations of a host conical state, but may lead to the formation of an exotic skyrmion order with pairs of merons being the core of the game. 
Generically, the fundamental insights provided by this work emphasize a three-dimensional character of skyrmion metamorphoses
and can also be extended to three-dimensional solitons, such as hopfions. 
}
\end{abstract}

         
\maketitle

\section{\label{intro} Introduction}
Multidimensional localized structures (topological defects or localized states) are the focus of research in many fields of modern physics \cite{30Schmeller,31Rossler}. Since the late 1960s, the problem of soliton-like solutions of non-linear field equations has been addressed
in condensed matter physics, biophysics, in particle and nuclear physics, in astrophysics, and cosmology \cite{32Rebbi}. 
From fundamental point of view, the interest in such solutions is related to the explanation of countable particles in continuous fields.
Within the structural theory the particle-like properties are ascribed to localized solutions of nonlinear field equations and the physical fields are described by the asymptotic behavior of corresponding solutions. 
Hobart and Derrick \cite{3334Hobart}, however, found with general arguments that multidimensional localized states are unstable
in many physical field models: inhomogeneous states may appear only as dynamic excitations, but static configurations collapse spontaneously into topological singularities. As a consequence, the solutions of corresponding non-linear field equations are restricted to one-dimensional solitons.
The instabilities of localized field configurations can be overcome, if the energy functionals contain, for example, contributions with higher-order spatial derivatives. 
This original idea of Tony Skyrme in 1960s succeded in describing the nuclear particles as localized states \cite{35Skyrme}. And the name "skyrmion" after Skyrme implies these localized solutions.

From the other side, the instability of multidimensional localized states may be avoided in condensed matter systems with broken inversion symmetry, where chiral interactions (energy terms linear with
respect to spatial derivatives of order parameters) play the crucial role in their stability.
In condensed matter physics chiral interactions arise due to structural handedness. Particularly, in magnetic non-centrosymmetric crystals chiral asymmetry of exchange interactions originates from relativistic Dzyaloshinskii-Moriya coupling \cite{Dz58,Moriya}. 
Phenomenologically, the Dzyaloshinskii-Moriya interaction (DMI) is expressed as the first derivatives of the magnetization  $\textbf{M}$ with respect to the spatial coordinates, the so called Lifshitz invariants (LI):
\begin{equation}
\mathcal{L}^{(k)}_{i,j} = M_i \partial M_j/\partial x_k - M_j \partial M_i/\partial x_k.
\label{LI}
\end{equation}
%
DMI can stabilize  one-dimensional (spirals or helices \cite{Dz64}) and two-dimensional (skyrmions) 
states  with a fixed  sense of the magnetization rotation. 
In particular, solutions for chiral magnetic skyrmions as static states localized in two dimensions have been derived first in 1989 \cite{JETP89}. It was shown that these topological solitonic field configurations exist in magnetic systems for all non-centrosymmetric crystallographic classes that allow Lifshitz invariants (\ref{LI}) in their magnetic free energy \cite{JETP89,Bogdanov94}.

Chiral interactions having the same functional  form as (\ref{LI}) may appear also in many other systems: in ferroelectrics with a non-centrosymmetric parent paraelectric phase, non-centrosymmetric superconductors, multiferroics \cite{Bogdanov01,Bode07,Wright},  or even in metallic supercooled liquids and glasses \cite{Sethna}. Localized states in these systems are also named skyrmions by analogy with the Skyrme model for mesons and baryons \cite{Skyrme}.

Chiral liquid crystals (CLC) are considered as  ideal model systems for probing behavior of different modulated structures on the mesoscopic scale \cite{Ackerman}. In these systems, a surprisingly large diversity of naturally occurring and laser-generated topologically nontrivial solitons with differently knotted nematic fields has been recently investigated \cite{Smalyukh10,Ackerman,Ackerman16}.
In particular, toron  represents a localized particle  consisting of two Bloch points at finite distance and a convex-shaped skyrmion stretching between them \cite{Ackerman,Inoue2018,toron1}. 
In chiral liquid crystals, the acentric shape of underlying molecules is at the heart of chiral effects.

In the last years, there is a renewed interest in studies of multidimensional topological solitons in chiral magnets and liquid crystals inspired by the discovery of two-dimensional magnetic skyrmions. 
Due to their nanometer size, topological protection and the ease with which they can be manipulated by electric currents, the magnetic skyrmions are considered as promising objects for the next-generation memory and logic devices. In particular, in the skyrmion racetrack \cite{Tomasello14,Muller17,Wang16} -- a prominent model for future information technology --  information flow is encoded in  isolated skyrmions (IS) \cite{Fert2013} moving within a narrow strip.
LC-skyrmions with a typical size of several micrometers can also be set in a low-voltage-driven motion with the precise control of both the direction and speed \cite{Ackerman17c}. 
The LC-skyrmions are confined in a glass cell with thickness comparable with the helicoidal pitch, an LC counterpart of a race-track memory. 
The most obvious application of LC skyrmions, however, is envisioned in various photonic crystal lattices and diffraction gratings, especially if one recalls others textures realized in liquid crystals and based on the different types of defects.

In thin layers of cubic helimagnets, skyrmions remain essentially 2D, but gain their stability due to the additional surface twists \cite{Rybakov2013,twists,twists1,twists2} involving the LIs  $\mathcal{L}^{(z)}_{x,y}$ with the magnetization derivative along $z$. 
These skyrmions were  first directly observed  in nanolayers of cubic helimagnets (Fe$_{0.5}$Co$_{0.5}$)Si \cite{yuFeCoSi} and FeGe \cite{YuFeGe} over a broad range of temperatures and magnetic fields.   
%
%
%
Topologically being elements of the 2nd homotopy group of spheres \cite{Smalyukh2020}, the two-dimensional magnetic skyrmions can be embedded in thin films or in the bulk of three-dimensional (3D) material systems as translationally invariant structures. However, the 3D configuration/physical space allows for more degrees of freedom in terms of orientation and spatial morphology of axes of these topological solitons, albeit the spectrum of different such possibilities is constrained by energetics of the system.    

Bulk magnetic systems represent a  truly 3D arena for skyrmion stabilisation and manipulation. 
In particular, bulk helimagnets enable propagation of skyrmion tubes either along or perpendicular to an applied magnetic field \cite{LBK,new} and thus may host rather complex topological structures. 
However until very recently skyrmions in these systems were found only in a small pocket of their temperature-magnetic field phase diagram,  just below the transition temperature $T_C$, the so-called A phase \cite{Kadowaki1982,Muehlbauer09}. This  restricted their potential as candidates for future spintronic devices. 
The recent discovery of low-temperature skyrmions in the bulk-insulating cubic helimagnet Cu$_2$OSeO$_3$  returns the focus of skyrmionic community to these 3D-embedded skyrmions \cite{Chacon2018,Bannenberg2019}. 
Moreover, the recent direct visualization of skyrmion clusters with mutually orthogonal orientations in CLC \cite{new} as well as the possibility to construct skyrmionic networks \cite{Vlasov}  strengthens this motivation. 
In this sense, skyrmionic superstructures constitute a novel concept of spintronic devices based on gapless skyrmion motion along each other \cite{Vlasov}.

\begin{figure*}
\includegraphics[width=1.99\columnwidth]{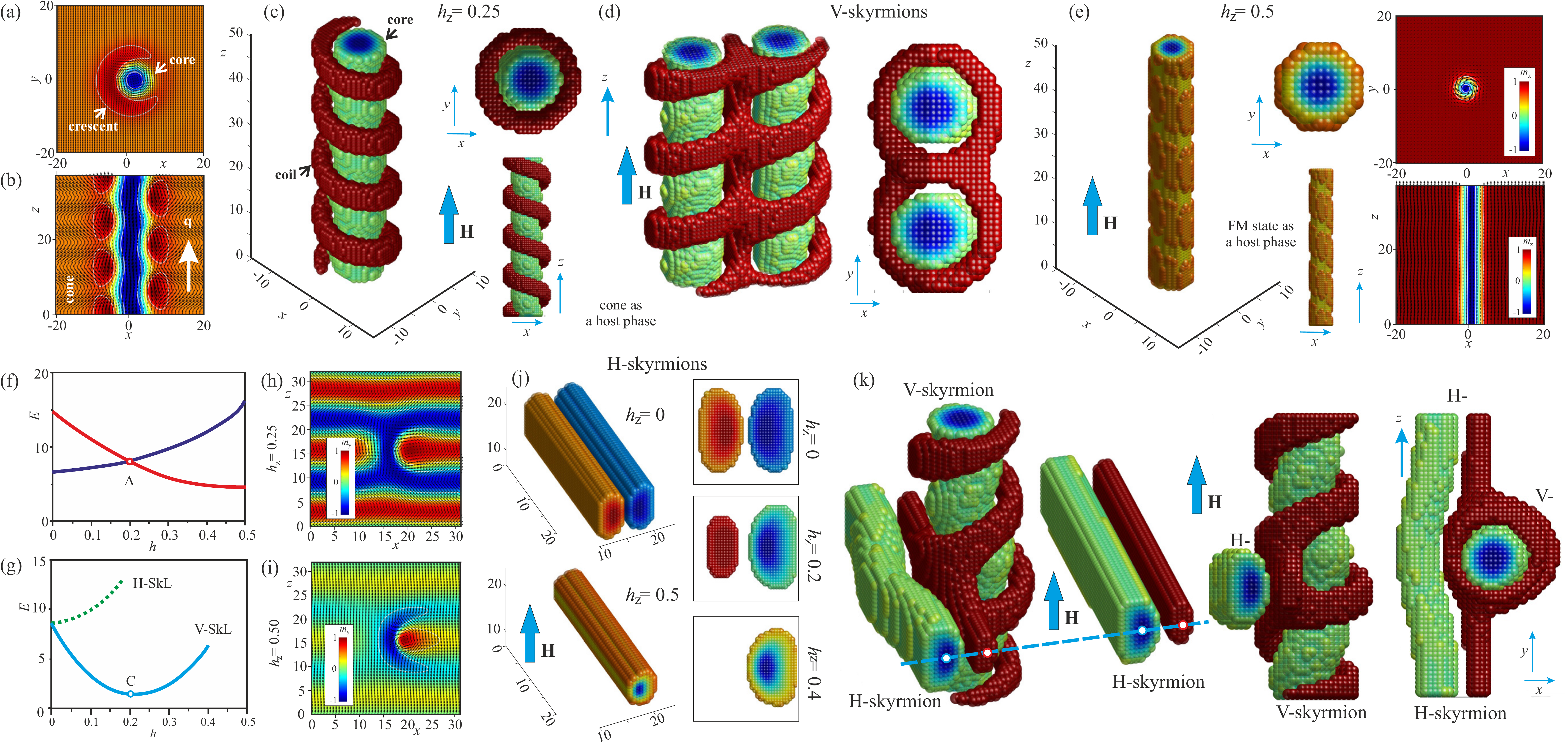}
\caption{(color online) Internal structure of attracting vertical (a)-(c) and horizontal (h), (j) skyrmions within the conical phase of bulk helimagnets and their field-driven transformation into repulsive particles within the homogeneous background (e), (i). The structure of isolated skyrmions is represented either as color plots of $m_z$ (a), (b) (or $m_y$ (h), (i)) components of the magnetization with arrows indicating projections of the magnetization onto the corresponding plane or by the three-dimensional  constructs with excluded spins of the conical phase (c), (j). Then, the cluster formation of vertical skyrmions (d) and mutually orthogonal skyrmions (k) is based on the latter picture: one simply assembles the skyrmions to achieve a compact defect-free configurations. 
Energy of V- and H-skyrmions above the host conical state in (e) indicates that the region of cluster formation embraces an intermediate field range below the cone saturation $h_c$ (point A in (e)). Energy densities of skyrmion lattices composed from the skyrmions of just one type indicate cogent preference of V-SkLs (see text for details).
\label{old}}
\end{figure*}

In the following we will unveil the principles of skyrmion meshing into a large diversity of extended three-dimensional skyrmionic networks and new phases in chiral magnets and liquid crystals. 
%
In the next section, we  introduce our phenomenological model and the algorithms used for our simulations. 
Within the continuous and discrete models under consideration, the applied magnetic field is the only control parameter.  Thus, we introduce the different critical values of the field to indicate different regimes of the host spiral states that embrace isolated skyrmions. 
In particular, the field-dependent conical phase may favor a particular type of isolated skyrmions and underlies the strength of the skyrmion coupling and thus the mutual distances between them. 

In Sec. \ref{reminder}, we examine the internal structure of IS tubes propagating correspondingly along and perpendicular to the wave vector of the conical state.
The constructed 3D models provide a straightforward explanation of  the attractive nature of skyrmion-skyrmion interaction. 
%
In Sec. \ref{target}, we consider the winding of  a horizontal (H) skyrmion around another vertical (V) one driven by the strong interaction energy between the two.
Depending on the cluster configuration, this may lead to a family of so called target-skyrmions with  alternating  topological charge ($Q=0, \,1$) or to complex multiple-$Q$ target states.
We also indicate that the wiggling instability of H-skyrmions may lead to spring-like states that may be seen as a missing intermediate element between H- and V-skyrmions (Sec. \ref{spring}). 
Finally, we consider the problem of skyrmion stability within the chiral soliton lattice for a field perpendicular to the cone wave vector (Sec. \ref{Hx}). Whenever possible, we discuss connections of our theoretical results to the experimental findings in different condensed-matter systems.

\section{\label{model} Phenomenological model }
In  modern skyrmionics, the phenomenological functionals introduced by Dzyaloshinskii in their most basic form, see Eq.  \ref{density},  are being used as main models for the interpretation of experimental results in different classes of non-centrosymmetric magnetic materials and beyond as a general foundation for the study of modulated phases \cite{NT,roadmap,ourreviews}. 
(see bibliography in Refs. \onlinecite{Levanyuk,Izyumov84,Cummins,Bogdanov94,Bogdanov02k} for earlier reviews). 
%
It is remarkable that all  effects on the 2D and 3D scales can be treated by the same basic model. To make the model more quantitative and directly related to a particular experiment, one usually supplements it with cubic and exchange anisotropies.

This standard model for magnetic states in bulk cubic non-centrosymmetric ferromagnets is based on the energy density functional \cite{Dz64,Bak80}
\begin{equation}
w =A\,(\mathbf{grad}\,\mathbf{m})^2 + D\,\mathbf{m}\cdot \mathrm{rot}\,\mathbf{m} -\mu_0 \,M  \mathbf{m} \cdot \mathbf{H}. 
\label{density}
\end{equation}
which includes the exchange stiffness and  the Dzyaloshinskii-Moriya constants, $A$ and $D$ respectively, as well as  the Zeeman energy. These are the principal interactions needed to stabilize all modulated states under scrutiny, e.g., one-dimensional (1D) spiral states and two-dimensional (2D) chiral skyrmions (which may become truly 3D with the structure modulations along the complementary third coordinate). 
$\mathbf{m}= (\sin\theta\cos\psi;\sin\theta\sin\psi;\cos\theta)$  is the unity vector along the magnetization vector  $\mathbf{M} = \mathbf{m} M$. $\mathbf{H}$ is a  magnetic field, which will be applied both along $z$ and in the transverse plane. 

For the forthcoming calculations, we use  non-dimensional variables defined in a consistent way with Ref. \onlinecite{Bogdanov94}. 
The lengths are expressed in units of $L_D=A/D$, i.e. the length scales are related to the period of the spiral state in zero field $p_0=4\pi L_D$, and reflect the fact that the ground state of the system in the form of a single-harmonic mode is yielded as a result of the competition between the counter-acting exchange and DM interactions in  Eq. (\ref{density}). Thus $L_D$ introduces a fundamental length characterizing the magnitude of chiral modulations in non-centrosymmetric magnets. 
$h = H/H_D$, where $H_D = D^2/(A M)$, is the reduced magnitude of the applied magnetic-field.

In the present manuscript as a host for isolated skyrmions, we consider spiral states with the wave vector $\mathbf{q}$ exclusively along the $z$ axis. 
For a magnetic field co-aligned with $\mathbf{q}$, a  conical spiral retains its single-harmonic character with:
\begin{equation}
\psi = \frac{2\pi z}{p_0},\quad \cos \theta = \frac{2 |\mathbf{H}|}{H_D},
\label{cone1}
\end{equation}
where $\psi$ is  the azimuthal angle. In such a helix the magnetization component along the applied field has a fixed value $M_{\bot} = M \cos \theta = 2MH/H_D$ and the magnetization vector $\mathbf{M}$ rotates within a cone surface. 
In this case, the conical state combines properties of the homogeneous state and the flat spiral as a compromise between Zeeman and DM energies and represents the global minimum of the functional (\ref{density}).
The critical value 
\begin{equation}
h_c=0.5 
\label{hc}
\end{equation}
marks the saturation field of the cone phase into the homogeneous state.

If a magnetic field is directed along the $x$ or $y$-axes, i.e., in the case where the propagation vector $\mathbf{q}$ of a spiral state is perpendicular to an applied magnetic field, the spiral state transforms into a chiral soliton  lattice (CSL) \cite{Dz64,Togawa}. 
While the polar angle retains its constant value $\theta=\pi/2$, the azimuthal angle $\psi$ is expressed as a set of elliptical functions and describes a gradual expansion of the CSL period with increased magnetic field. In a critical magnetic field 
\begin{equation}
h_h=\pi^2/8=0.30843
\label{hh}
\end{equation}
the CSL infinitely expands and transforms into a system of isolated non-interacting $2\pi$-domain walls (kinks) separating domains with the magnetization along the applied field. 
The CSL is a metastable solution of the functional (\ref{density}), since the conical state with $\mathbf{q}$-vector co-aligned with the field is the global minimum. However, it is instructive to consider isolated skyrmions also within this metastable CSL-state, since the $\mathbf{q}$-vector may be pinned by additional anisotropic contributions and may retain its perpendicular orientation with respect to the field.
Such a behavior was recently observed, e.g., in the polar magnetic semiconductor GaV$_4$S$_8$. The cycloid states were found to point along the $\langle 110  \rangle$ directions irrespective of the field orientation \cite{Kezsmarki15}. For a magnetic field perpendicular to the $\mathbf{q}$-vectors, the conical state becomes the global minimum of the system. A magnetic field in the plane of the magnetization rotation, however,  leads to a CSL resulting to the reported anomalies of the phase diagrams \cite{Geirhos}.

\begin{figure*}
\includegraphics[width=1.7\columnwidth]{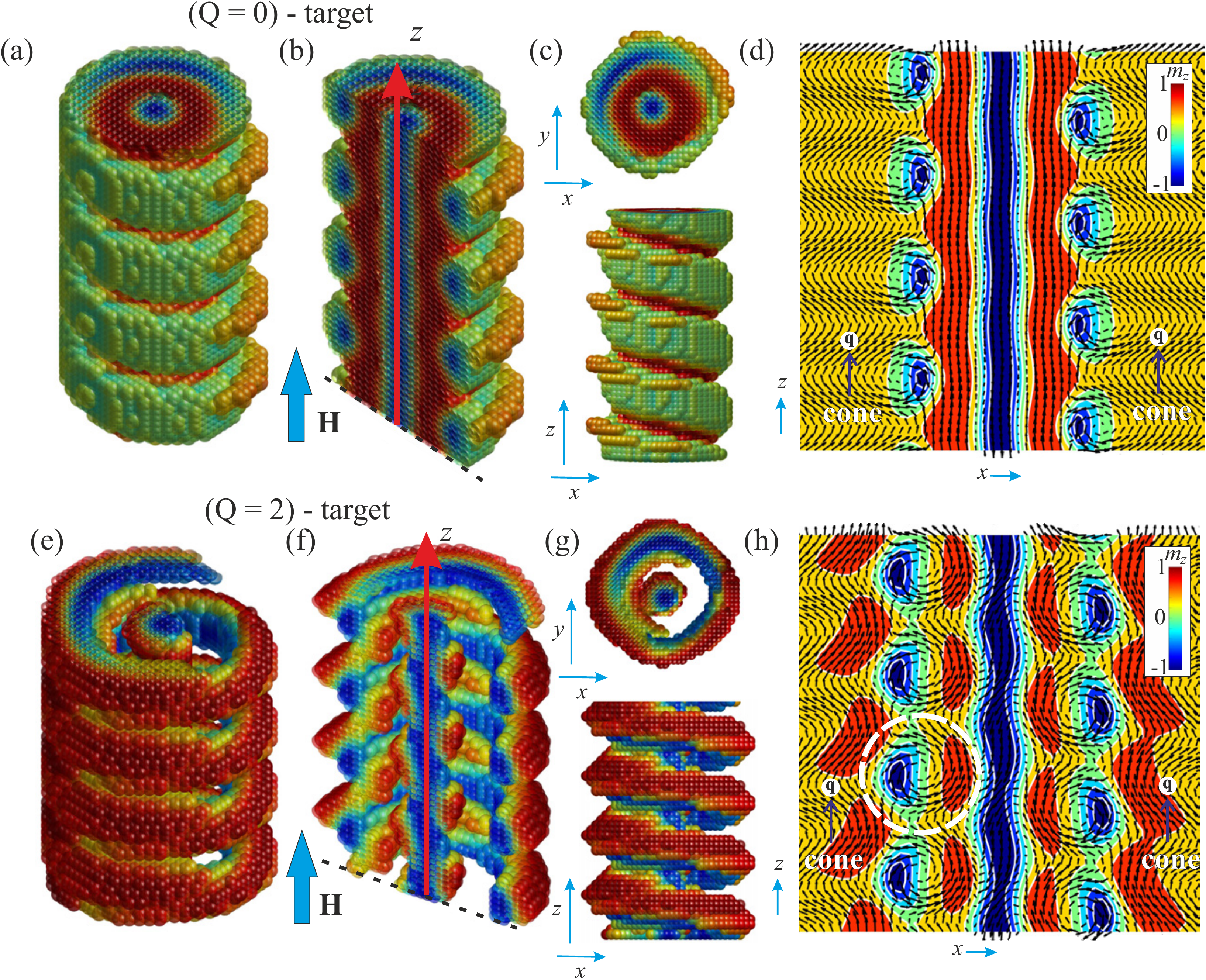}
\caption{(color online) Internal structure of target-skyrmions with either an alternating topological charge ($Q=0;\,1$, (a)-(d)) or the multiple $Q$ ($Q=2$, (i)-(h)). Target-skyrmions originate as a result of swirling of H-skyrmions around a V-skyrmion (see text for details)  by the tendency to reduce the interaction energy. (b) and (f) (as well as (d) and (h)) exhibit dissected targets for better examination of their core structure. 
\label{targets}}
\end{figure*}

\section{\label{reminder}  Skyrmions within the conical state}

In this section we consider the internal structure and characteristic features of  skyrmions within the conical state for $\mathbf{h}||z$. 
In cubic helimagnets ISs may  orient themselves either along or perpendicular to the field. 

The internal spin pattern of  V-IS with their axes along the wave vector of the conical phase is depicted in Fig. \ref{old} (a),  (b). 
In the field range $[0,\,  h_c]$, the magnetization distribution in a cross-section $xy$ splits into the central core region that nearly preserves the axial symmetry (Fig. \ref{old} (a)) and the domain-wall region, connecting the core with the embedding conical state. This part of the skyrmion cross-section is \textit{asymmetric} and acquires a crescent-like shape, which undergoes additional screw-like modulation along $z$ axis, matching the rotating magnetization of the conical phase (Fig. \ref{old} (b)).

A convenient way to depict these skyrmions, which has been proven to be particularly illustrative in addressing the character of skyrmion-skyrmion interaction, is as follows (Fig. \ref{old} (c)): we extract the spins  corresponding to the conical phase and then plot the remaining spins as spheres colored according to their $m_z$-component. 
In this way, all  intricate details of the internal structure are explicitly revealed, which spares the difficulties to plot skyrmion crosscuts along the different directions. 
In such a fashion, V-skyrmions are composed of a cylinder-like (blue) core centered around the magnetization opposite to the field and a (red) coil with the magnetization along the field (Fig. \ref{old} (c)). 
Current experimental endeavors are particularly focused on unveiling the three-dimensional spin texture of skyrmion tubes \cite{Damien,twists1,twists2,Birch:2020}.

\begin{figure*}
\includegraphics[width=1.99\columnwidth]{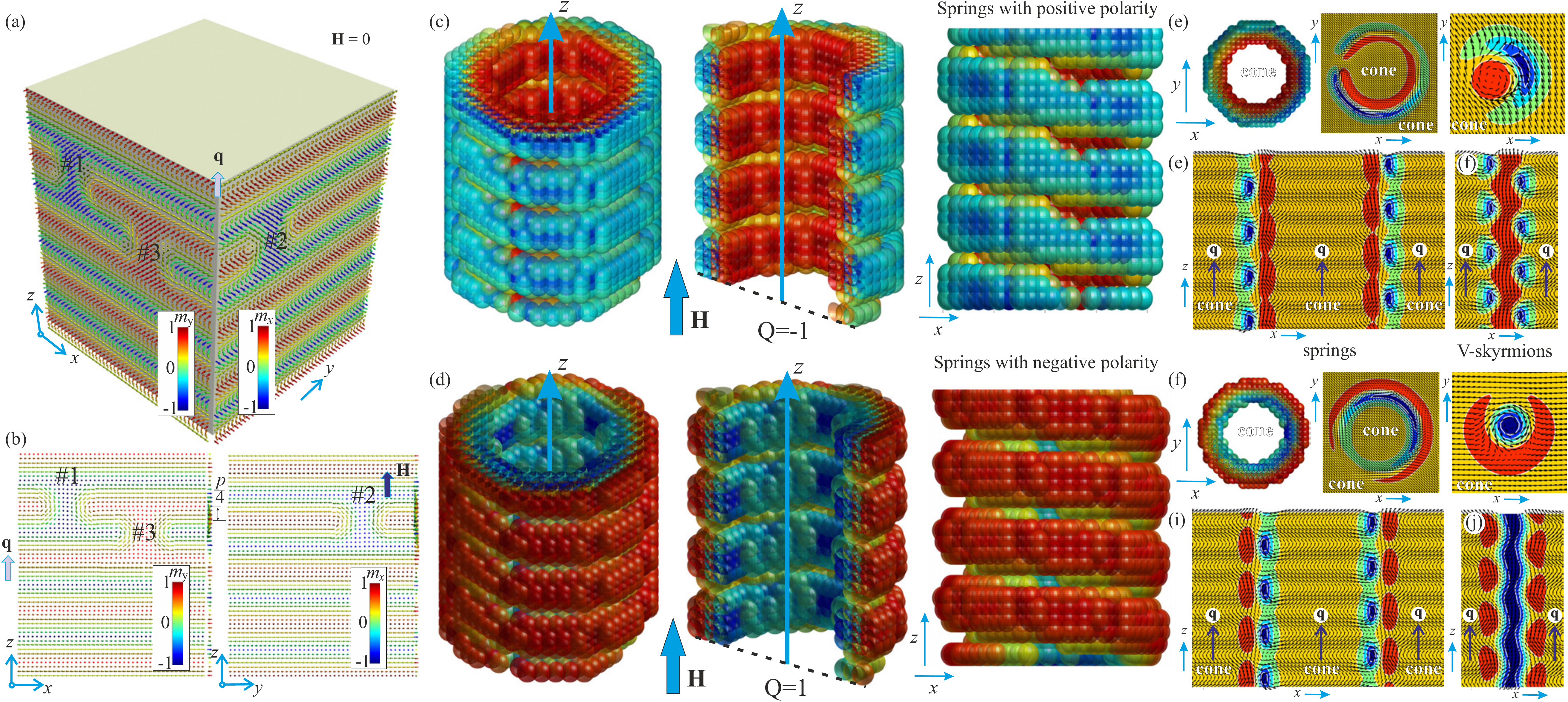}
\caption{(color online) Internal structure of spring-like states obtained by swirling of H-skyrmions around $z$ axis and guided by the small interaction energy. Depending on the interior, the springs acquire either the positive  (c) or the negative (d) polarity, as well as the negative and the positive topological charges. This effect is based on the varying altitude along $z$, if an H-skyrmion undergoes wiggling instability (a), (b) (see text for details). Since within the model (\ref{density}) the energy of the springs is positive, they contract to a vertical skyrmion with the corresponding polarity (e), (f). 
\label{springs}}
\end{figure*}    

The cluster formation of V-skyrmions may be envisioned  as a process of zipping  loops: a coil of one V-skyrmion penetrates the voids between the coils of another one (Fig. \ref{old} (d)). By this, the compact skyrmion pair recreates a fragment of a SkL that within the model (\ref{density}) is a metastable state: the magnetization on the way from the center of one skyrmion to the center of another rotates as in ordinary axisymmetric skyrmions.

Attraction of V-IS mediated by the conical phase was considered theoretically  in Refs. \onlinecite{LeonovJPCM16,LeonovAPL16,Ackerman17c,attraction1,attraction2}. Experimentally, clusters of such skyrmions have been observed in thin (70 nm) single-crystal samples of Cu$_2$OSeO$_3$ taken using transmission electron microscopy \cite{Loudon18} and in nematic fluids, where they were shown to also form skyrmion chains \cite{Ackerman17c}. 
In the following, we  refer to the experiments  on both the chiral magnets and chiral liquid crystals that support  these theoretical results. 
%

For $h>h_c$ with the onset of the homogeneous state, V-ISs loose their coils and  acquire an axisymmetric structure localized in nanoscale cylindrical regions (Fig. \ref{old} (e)) thus representing ensembles of weakly repulsive particles. Repulsion of conventional axisymmetric V-skyrmions within the saturated state was investigated in Ref. \onlinecite{LeonovNJP16} and it was shown that the  region where  V-skyrmions persist at high fields is defined by their collapse. 
The field-driven evolution of V-skyrmions is driven by  their decreasing energy from the most distorted state within the conical phase at $h=0$ to the axially symmetric particles at higher fields (red curve in Fig. \ref{old} (f)).
%

\begin{figure*}
\includegraphics[width=1.99\columnwidth]{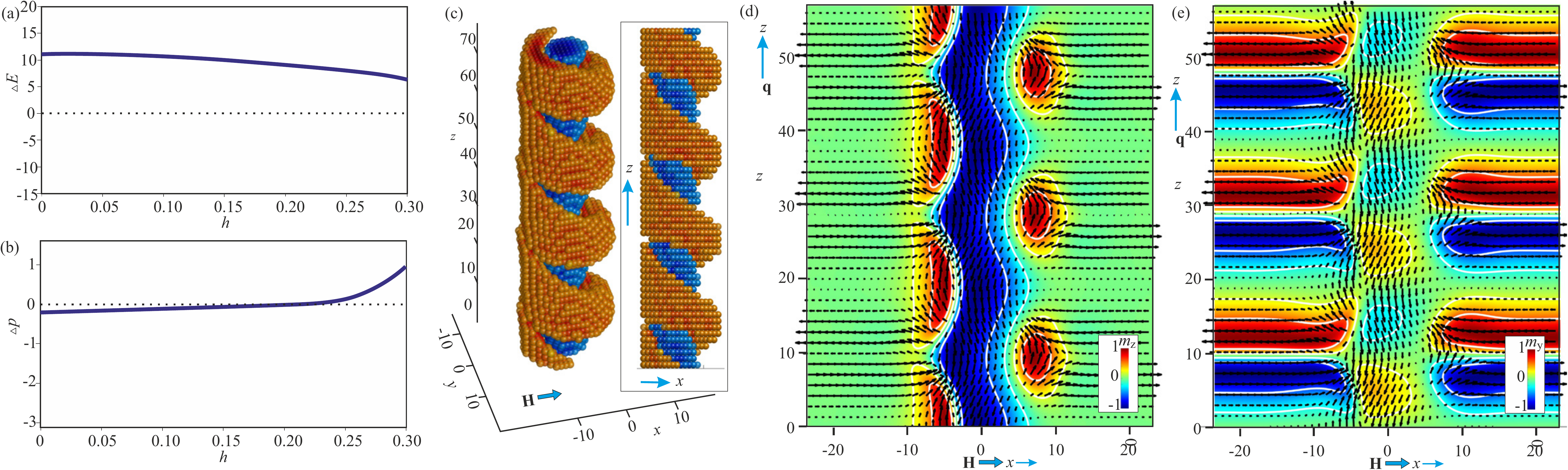}
\caption{(color online) Field-driven deformation of a V-skyrmion within the chiral soliton lattice for the in-plane direction of the magnetic field,  $\mathbf{H}\perp z$ (see text for details). Color plots exhibit $m_z$ (d) and $m_y$ (e) components of the magnetization with the arrows showing projection of the magnetization onto the $xz$ plane. Alongside with the 3D skyrmion model (c), the pictures demonstrate broadening of skyrmion loops along the field and narrowing otherwise, which is consistent with the spiral behavior in the field perpendicular to its wave vector. 
The eigen-energy of a V-skyrmion remains positive with respect to the surrounding spiral phase (a). Still, an isolated skyrmion intervenes into a field dependence of a spiral period (b). 
At a low magnetic field, a V-skyrmion slightly expands the spiral, whereas for the field near the saturation (\ref{hh}), it slows down a process of spiral expansion. Here $\Delta p$ is the difference between the periods of spiral states without and with a V-skyrmion. 
\label{vertX}}
\end{figure*}

Horizontal skyrmions perfectly blend into the spiral state and for $h = 0$ they represent a pair of merons with equally
distributed topological charge $Q = 1/2$. Figs. \ref{old} (h), (i) show the evolution of such a meron pair into a distorted isolated skyrmion. 
Red and blue areas in Fig. \ref{old} (h), (i) correspond to positive and negative values of the $m_y$-component. 
Following the same paradigm of excluded cone spins, H-skyrmions are displayed in Fig. \ref{old} (j) as two parallel distorted cylinders centered around the core (blue) and the transitional region (red) with the negative and the positive $m_z$ component of the magnetization, correspondingly. For consistency, we run H-skyrmions along the $y$ axis and keep in mind that these may exist  with positive and negative polarities, i.e. directions of the magnetization in the region between two cylinders.

The evolution of H-skyrmions with  increasing magnetic-field is shown in Fig. \ref{old} (j).  At higher fields, they have non-axisymmetric structures on the  $xz$ plane and within a given accuracy of the excluding spin procedure, the "red" counterpart of H-skyrmions disappears altogether within the ferromagnetic surrounding.
Thus, the field-driven transformation of H-skyrmions leads to a clear energetic disadvantage at higher fields (blue curve in Fig. \ref{old} (f)).  
Once embedded into the conical phase,  H-skyrmions develop an attracting interaction, which becomes repulsive within the homogeneous state. According
to Refs. \onlinecite{Vlasov,Muller}, two coupled H-skyrmions are energetically more favorable than two distant ones. However,  the energy difference between these two configurations is almost negligible. Therefore, these clusters are less coupled than the corresponding clusters of V-skyrmions, which have a much higher dissociation energy, up to $30\%$, in comparison to the  energy of  V-ISs \cite{LeonovJPCM16}.

A consequence of the above mentioned considerations is that the crossover between the two types of ISs (dubbed a skyrmion flop transition in Ref. \onlinecite{Vlasov} or toggle-switch-like crossover in Ref. \onlinecite{LBK}) takes place for an intermediate value of the field $h \approx 0.2$ (Fig. \ref{old} (f)). Furthermore, there is an obvious advantage of V-skyrmion clustering to SkLs composed of skyrmions of the same type  (Fig. \ref{old} g)).

Cluster formation of mutually orthogonal skyrmion tubes (Fig. \ref{old} (k)) is also  possible, as shown  in Ref. \onlinecite{new}. The energy gain due to the clustering of H- and V-skyrmions was shown to reach the same high value of  $30\%$ in comparison to  the energy of  uncoupled ISs.
For coupled H- and V-skyrmions one may distinguish two cluster configurations (Fig. \ref{old} (k)). Indeed, the transient red region of one H-skyrmions may either slide in-between the coils of V-skyrmion or just touch its coil if located on the other side.
Here we notice, however, that within the basic model (\ref{density}) both skyrmion varieties have a positive energy with respect to the host phase. 

Both regimes of skyrmion interaction were directly visualized in the chiral nematic LC mixtures \cite{new}. 
In these experiments, the V- and H-skyrmions were controllably "drawn" into the LC cells using optical tweezers. 
Additionally, they were manipulated by the focused laser beams and spatially translated within the sample plane.
Because of the chiral LCs preference to twist, the skyrmions were proved to be topologically stable excitations in the conical background.

\begin{figure*}
\includegraphics[width=1.99\columnwidth]{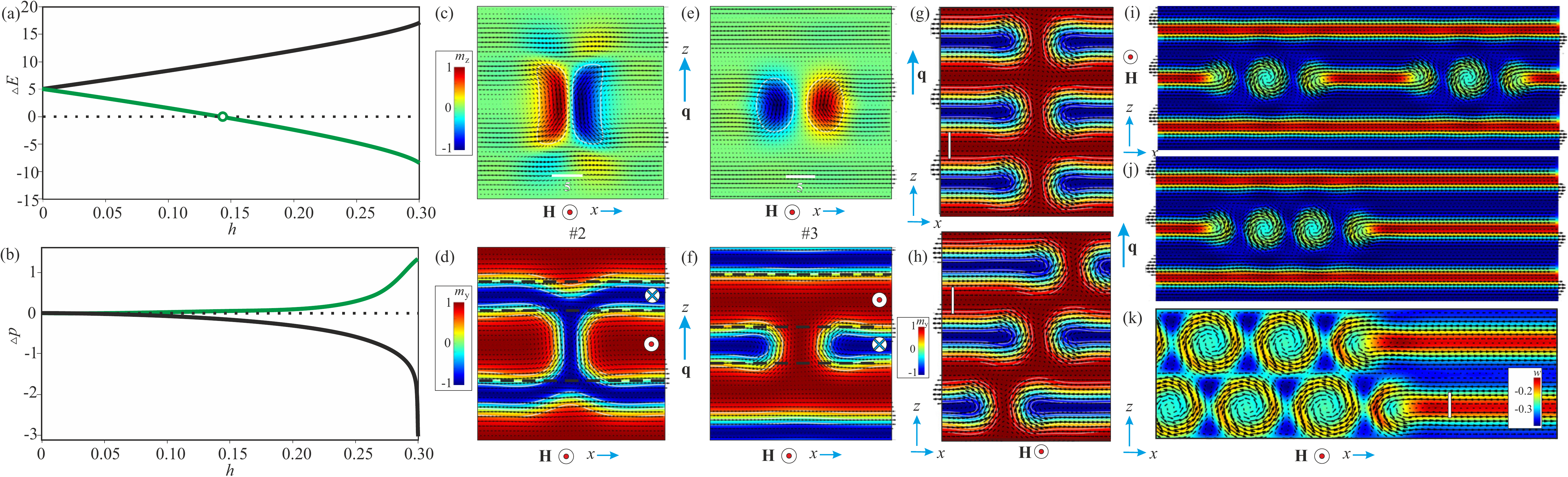}
\caption{(color online) Field-driven evolution of H-skyrmions for the field coaligned with their axes.  Both types of H-skyrmions are exhibited either as color plots of $m_z$ magnetization components (c), (e) or $m_y$ ones (d), (f) and represent cuts of a spiral state in the form of two merons. 
%
%
Whereas a pair of merons with the negative polarity becomes favorite within the spiral state, merons with the positive polarity maintain their positive energy (a).
In the former state, an avalanche-like formation of merons eventually results in a hexagonal skyrmion lattice (k): first of all, a hexagonal-like arrangement of merons (h) is energetically more favorable as compared with the square-like arrangement (g); then, configurations with the skyrmions gathered together (j) are preferred over their dispersed arrangements (i). 
(b) provides an information on the distortions of the equilibrium spiral period made by both types of H-skyrmions.  
\label{horY}}
\end{figure*}

In chiral cubic helimagnets magnets, however, clusters of mutually orthogonal skyrmions may account for the anomalies within the A-phases provided 
the ISs have a negative energy  with respect to the surrounding conical state. 
This would underlie the nucleation processes of skyrmion filaments 
via torons \cite{Inoue2018,toron1}. 
In these systems the problem of V-skyrmion stability is fully resolved in contrast to that of  H-skyrmion stability, which is still unknown. 
Indeed, the stabilization/nucleation mechanism of V-skyrmions is based on the deformations, imposed by small anisotropic contributions, to their main competitor -- the conical phase -- rather than on  themselves. 
In particular, uniaxial anisotropy of the easy-axis type \cite{Butenko10}, which does not affect the ideal single-harmonic type of the magnetization rotation in the conical spiral but just leads to the gradual closing of the cone, grants the thermodynamical stability of the V-SkL in a broad region of theoretical and experimental phase diagrams \cite{uniaxial1,uniaxial2,uniaxial3}.
Cubic anisotropy also stabilizes V-SkLs but for specific directions of the applied magnetic field \cite{Leonov2020,Bannenberg2019,cubic1,cubic2}. 
This anisotropy deforms the ideal conical configuration as the magnetization tends to deviate from the ideal conical surface trying to embrace the easy axes and to avoid the hard directions. 

In the case of H-skyrmions, however,  any manipulation  of the conical phase  does not necessarily improve the skyrmion's stability. On the contrary, it may even increase  the eigen-energy of H-skyrmions.  

In case there is a stabilization mechanism suitable for both skyrmion species, H-skyrmions will lower their energy around the most undistorted state at $h_z=0$ whereas V-skyrmions will be stabilized at rather high fields, These two species would thus be responsible correspondingly for zero-field and A-phase precursor effects  in cubic helimagnets \cite{MnSi,FeGe}. 
In the following, we do not address the question of skyrmion stability but concentrate on already existing/created metastable H- and V-skyrmions as was done, e.g., in the experiments on chiral nematics \cite{new}.



\section{\label{target} Target-skyrmions as a result of skyrmion addition and/or subtraction}

Here we consider the attraction between  mutually orthogonal skyrmions, which  opens up a route for skyrmion addition and/or subtraction. 
We first treat the case of an H-skyrmion that runs parallel to the $y$ axis in Fig. \ref{old} (k) and is ziplocked  with a V-skyrmion.
Guided by the large negative interaction energy, the H-skyrmion may start to wind around the V-skyrmion penetrating the space between the coils and circling along $z$. This process leads to further energy reduction with respect to the configuration shown in Fig. \ref{old} (k). 
The obtained target-skyrmion bears a total topological charge of $Q=0$ and is therefore the result of a topological charge subtraction between the H- and a V-skyrmions.
Alternatively, such a composite object can be regarded as a V-skyrmion with $Q=0$, in which an H-skyrmion contributed to its structure.
%
%

In the transverse plane $xy$, such a  target-skyrmion  consists  of  a doubly-twisted core  surrounded  by a number of concentric helicoidal undulations (Fig. \ref{targets} (a)-(d)).
Depending on the number of these concentric  helical  stripes, the topological charge alternates and has the value either 0 (odd number of helical undulations) or 1 (even number). 
Such target-skyrmions have been recently thereotically predicted \cite{target1} and subsequently experimentally observed in magnetic nanodiscs/nanowires \cite{target2,target3}. They may be energetically favoured  by both the magnetostatic energy demanding a flux-closed states and by the so-called edge states that supply targets with additional negative energy.

Surprisingly, a target-skyrmion with the topological charge $Q=2$ is also achievable if an H-skyrmion has its blue part touching the red coil of a V-skyrmion (Fig. \ref{old} (k)). In this case the H-skyrmion  will circle around the V-skyrmion, along the $z$ axis, and the two topological charges will be added to the total charge of a target-skyrmion rather than subtract from each other. 
The composite structure of such a state is illustrated in Fig. \ref{targets} (e)-(h). The $(Q=2)$-target is separated by some potential barrier from the $(Q=1)$-V-skyrmion, since a part of the structure (encircled by a white dashed line in Fig. \ref{targets} (h)) may annihilate.

Skyrmions with $Q=2$ are usually obtained via the   in-plane magnetization rotation\cite{Q2,Q22} : in such skyrmions, the polar angle varies from $\pi$ to 0, but the azimuthal angle becomes:  $\psi=Q\varphi+\gamma$. This in general leads to  multiple-$Q$ skyrmions \cite{Q222}, which are fundamentally different from the ones considered in Fig. \ref{targets} (e). Another approach of obtaining high-charge skyrmions involves construction of  so-called "skyrmion bags" \cite{Foster2019}, which are translationally invariant along the skyrmion bag's tube and also fundamentally different from the ones considered in Fig. \ref{targets} (e), even if they can also be realized in chiral magnets and liquid crystals. 

In general, by winding H-skyrmions around V-skyrmions one may achieve  a family of target-skyrmions with either $Q=0$ or $Q=1$, depending on the  energetically stable cluster configuration. Thus, in order to create a target-skyrmion with $Q=1$ out of that in Fig. \ref{targets} (a), one should slide an H-skyrmion with its blue part between the blue coils of a current target-skyrmion.  
On the other hand, in order  to create a target-skyrmion with $Q=3$ out of that in Fig. \ref{targets} (e), the blue part of an H-skyrmion shoudl touch a red coil of a current target-skyrmion.
%

\section{\label{spring} Skyrmion springs with positive and negative polarities}

The wiggle instability of H-skyrmions may also be guided by the much smaller interaction energy that underlies the attraction between them \cite{Vlasov,Muller}. 
Also provided that 
 H-ISs acquire negative energy, their bending would serve the tendency to occupy the whole space, thus being an alternative mechanism to skyrmion condensation into a skyrmion lattice.

However, if H-skyrmions bend, they change their altitude along $z$. Indeed, in Fig. \ref{springs} (a) two straight H-skyrmions  running along $x$ and $y$ axes (marked by numbers $\#1$ and $\#2$) differ by the value $ossibly/4$ of their relative  $z$ coordinate. Another $\pi/2$-rotation results in an H-skyrmion running opposite to $y$ axis (marked by $\#3$)  and thus having an opposite polarity to that along the $y$ axis. 
The full rotation by $2\pi$ leads to two skyrmions stacked above each other at the distance $p$. 
The  sense of H-skyrmion circling is fully defined by the encompassing conical phase (counterclockwise in the present case) and is therefore unique.
%

\begin{figure*}
\includegraphics[width=1.99\columnwidth]{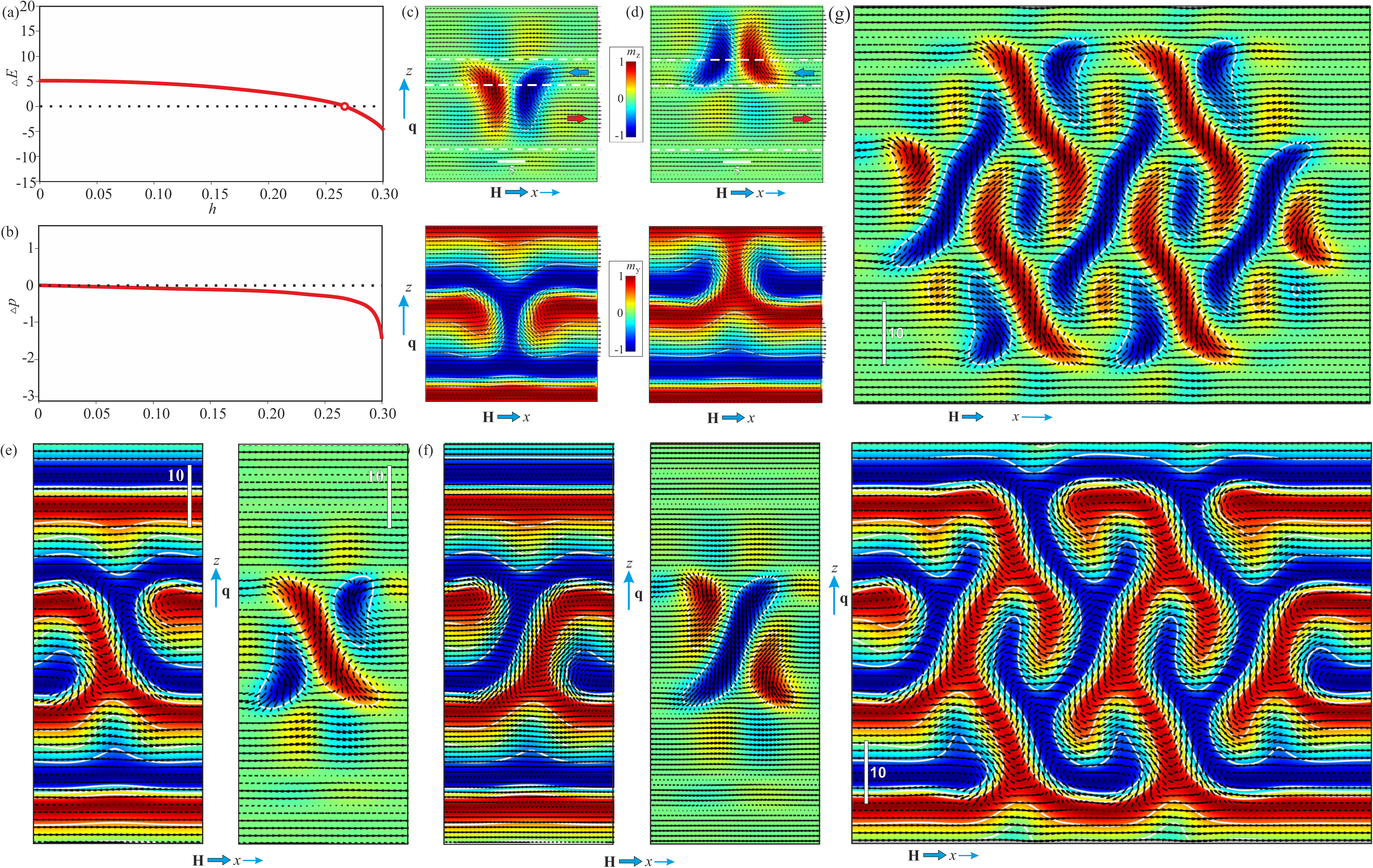}
\caption{(color online) Field-driven evolution of H-skyrmions for the field perpendicular to their axes.  Irrespective of their polarity, H-skyrmions bear the same energy over the host CSL state (red curve in (a)). The structure of H-skyrmions is shown by the color plots of $m_z$  and $m_y$ components of the magnetization (c), (d). The former graphical representation appears to be more instructive in elucidating the principles of meshing the H-skyrmions into entangled patterns (g). 
First of all, we notice that the energy of H-skyrmions becomes negative in the vicinity of $h_h$ (a) what favors spiral cuts. Then, the structural cross-like elements (e), (f), obtained as intersections of two H-skyrmions with opposite polarities, may become compositional units of skyrmions superstructures.  One of the examples that mixes "crosses" in an involved way is shown in (g) and has smaller energy as compared with the host spiral state. (b) provides an information on $\Delta p$.  
\label{horX}}
\end{figure*}

Wiggling of H-skyrmiosn  may also give rise to exotic textures in the form of skyrmionic springs, which exist in two varieties with the positive (red) and/or the negative (blue) part facing the interior (Fig. \ref{springs} (c) - (f)). 
%
%
Since within the basic model (\ref{density}) such springs possess  positive energy over the conical phase, they gradually shrink and in this way  transform into corresponding V-skyrmions with a positive or negative topological charge depending on their polarity (Fig. \ref{springs} (d), (f)).
In this sense, such springs can be considered as an intermediate link between two skyrmion varieties.  
In general,  springs with  variable radius could also be envisioned, but in this case the interaction energy is not fully minimized, since the overlap region of H-skyrmions diminishes. 
We also notice that even one loop of a spring contracted to a patch of a V-skyrmion (which then represents a toron \cite{Inoue2018,toron1}) provides a new mechanism of skyrmion nucleation, i.e., looping H-skyrmions gives rise to torons, which subsequently may be instigated to stretch along the $z$ axis owing to, e.g., cubic anisotropy.

The considered 3D springs with a constant tube radius along $z$ 
can be considered as relatives of 2D Archimede's spirals. In chiral LC, these spirals are produced by the transverse drift of fingers of the second type \cite{Oswald}. Since the fingers are confined within the thin film by the surface anchoring, their wiggling instability results in spiraling with the same altitude along $z$ and thus leads to a number of spiral turns. 

Beyong chiral  LCs, such  Archimede's spirals may exist in garnet ferrite films \cite{snail1,snail2}, where they are surrounded either by the labyrinth domain structure or by the bubble magnetic domains and are stabilized by the dipole-dipole interactions. By turn, the bending of stripe domains in garnet-ferrites reflects the geometry of easy anisotropy axes or their projections onto the basal plane. 
The structure of Archimede's spirals was also addressed numerically in chiral helimagnets within two host phases -- spirals and SkL \cite{snail3}. 
Thus in some sense, the considered 3D springs (Fig. \ref{springs}) may be regarded as Archimede's spirals escaping into the third dimension, but also driven by the mutual attraction between adjacent loops.  
Interestingly, geometrically more complex structures with local fragments resembling H-skyrmions can yield  a hopfion of the so-called "heliknoton" type, a topological solitons described by the 3rd homotopy group elements as topological invariants \cite{hopfion}.

\section{\label{Hx} Isolated skyrmions within the metastable spiral states for an in-plane magnetic field}

In the following, we examine the structure of H- and V-skyrmions within the spiral state $\mathbf{q}||z$ for an in-plane direction of the field $\mathbf{H}\perp z$. 
In this case, the complex three-dimensional internal structure of magnetic ISs and the character of the IS-IS interaction are imposed
by a surrounding parental state as investigated in detail in Refs. \onlinecite{new,Vlasov}. 
On the other hand, however, ISs themselves distort the structure of their host and in this way they  influence its field-driven evolution.

V-skyrmions, which retain their positive energy even over a metastable spiral state (Fig. \ref{vertX} (a)), are found also to impede a gradual expansion of the helicoid period with the critical field $h_h$ (\ref{hh}) (Fig. \ref{vertX} (b)).
For low values of the field, V-skyrmions  stimulate  spirals to slightly expand in order to release their own structural distortions, which  lead to a high positive energy. 

In Fig. \ref{old} (f), the energy of a V-skyrmion in zero field was computed with the fixed period $p=4\pi L_D$ of a spiral state. However,  some energy reduction can be achieved for a larger spiral period (Fig. \ref{vertX} (b)). 
With an increasing magnetic field, however, a V-IS defers the spiral expansion, since this process would lead to the increase of its own energy.
As a result, the energy of an IS within the spiral state with a non-equilibrium period, could be reduced.
Thus, the internal structure of V-skyrmions shown in Fig. \ref{vertX} (c)-(e) is balanced by the spiral tendency to acquire an equilibrium period for a given value of the field and the skyrmions' tendency to reduce their length. 
The skyrmion coils are deformed according to expanding and shrinking parts of a spiral state aligned, correspondingly, along and opposite to the field.

Obviously, the structure of V-skyrmions does not depend on the direction of the field within the plane $xy$.
%
For H-skyrmions, however, the two cases with $\mathbf{H}||x$ and $\mathbf{H}||y$ should be treated separately.
First of all, the $H_y$-magnetic field lifts the energetic degeneracy of H-skyrmions with the positive and the negative polarities (Fig. \ref{horY} (a),  (b)). 
Indeed, the cuts of an expanding spiral state in the form of two merons may occur either in the "wide" or in the "narrow" part of a spiral period (Fig. \ref{horY} (c) - (f)) with the latter being energetically more favorable (Fig. \ref{horY} (a)). 
Moreover, the energy of this meron pair becomes negative with respect to the helical background at a critical field $h_y\approx 0.16$  what then gives start to the conventional helicoid-SkL first-order phase transition observed experimentally in thin-film helimagnets \cite{yuFeCoSi,YuFeGe}.
The hexagonal arrangement of meron pairs (Fig. \ref{horY} (h)) is more favorable as compared with the square one (Fig. \ref{horY} (g)).
Furthermore, a configuration with closely packed skyrmions and merons (Fig. \ref{horY} (j)) due to a mutual attraction is more favorable than any configuration with dispersed merons (Fig. \ref{horY} (i)). 
Eventually, such a process of meron formation \cite{Muller,Ezawa} and condensation culminates in the hexagonal arrangement of closely packed skyrmions within the skyrmion lattice phase (Fig. \ref{horY} (k)). 
Interestingly, the critical field of the SkL-helicoid phase transition equals $H\approx 0.1 H_D$ if based only on the direct comparison of corresponding energies of a helicoid and a hexagonal SkL \cite{Bogdanov94,Butenko10}, thus, showing a field gap for merons to become energetically advantageous and to initiate an "avalanche"-like phase transition. 
Since the period of a CSL diverges at the critical field $h_h$, whereas the period of the SkL -- at $h_{sk}\approx 0.4 H_D$, a pair of merons (green curve in Fig. \ref{horY} (b)) slows down a process of a spiral expansion (see Ref. \onlinecite{Butenko10} to compare equilibrium periods of spiral and SkL states within the model (\ref{density})).
H-skyrmions with the opposite polarity (Fig. \ref{horY} (c), (d)) remain metastable states in the whole field range with the monotonous energy increase. Such skyrmions thus stimulate the CSL expansion (Fig. \ref{horY} (b)). 

In an $H_x$-magnetic field, however, the degeneracy of H-skyrmions persists as the two skyrmion configurations with differently oblique positive and negative parts (Fig. \ref{horX} (c), (d))  
bear the same energy (Fig. \ref{horX} (a)). 
Surprisingly, the energy of H-skyrmions becomes negative near the critical field $h_h$ what may result in the same process of meron formation as described previously for $\mathbf{H}||x$. 
However, besides the obvious  square-like or hexagonal-like arrangement of merons, one may consister the emergence of  more entangled spin structures. 
As a basis for these configurations, we consider cross-like states obtained by the intersection of two H-skyrmions with opposite polarities and having  positive or  negative diagonals as their main elements (Fig. \ref{horX} (e), (f)). 
An example of the periodic dendritic patterns obtained by meshing these cross-like elements is shown in Fig. \ref{horX} (g) and has  negative energy as compared with the host spiral state.


\section{Conclusions}

To conclude, we introduce an approach for numerical analysis of 3D chiral skyrmions. By scanning the intricate structure of skyrmions and excluding the spins corresponding to the host conical phase, we obtain 
representative 3D models of horizontal and vertical skyrmions that not only unambiguously characterize the topology of these 3D particles, but also provide tools to envision the process of complex cluster formation.

All our calculations have been based on the phenomenological Dzyaloshinskii theory for chiral magnets. 
Remarkably, this basic model is able to address a range of phenomena rooted in the chiral DMI 
and is widely used for interpretation of experimental results in non-centrosymmetric magnetic materials, and beyond as a general foundation for the study of modulated phases and localized states. 
Moreover, due to the deep connection of phenomenological models for different condensed-matter systems, the results of numerical
simulations were shown to successfully address solitonic textures in other condensed-matter systems, in particular, in chiral liquid crystals.

Our work also unravels  a path that  allows to transform horizontal skyrmions into vertical ones with appropriate polarity, which is achieved via the wiggling instability of H-skyrmions.
We speculate that undulations of H-skyrmions may generate torons -- nuclei of V-skyrmions that are able to elongate in presence of stabilizing anisotropic interactions. 
So called spring-like states appear as ultimate states of such   H-skyrmions swirling  along the $z$ axis driven by attraction between adjacent loops. 
A vertical skyrmion,  lassoed by a horizontal one, also predefines a direction of their propagation and may lead to a family of target-skyrmions including those with a multiple topological charge.

Finally, we illustrate a reverse impact of isolated skyrmions on the structure of the host spiral states for the field perpendicular to the spiral wave vector.
Having a positive energy over the host state, skyrmions affect the process of a field-induced spiral expansion.
Acquiring a negative energy, skyrmions initiate a first-order phase transition into a corresponding skyrmion arrangement, which is not necessarily a hexagonal skyrmion lattice: exotic skyrmionic networks can also be envisioned.

\textit{Acknowledgements. }
The authors are grateful to Ulrich R\"o\ss ler 
for useful discussions. AOL thanks Ulrike Nitzsche for technical assistance.C.P acknowledges financial support from the Vrije FOM program "Skyrmionics". I.I.S. acknowledges support of the US National Science Foundation grant DMR-1810513.

\end{document}